\pdfoutput=1
\RequirePackage{fixltx2e}
\documentclass[11pt,a4paper]{article}
\usepackage[utf8]{inputenc}
\usepackage{jheppub}
\usepackage{stmaryrd}
\hypersetup{unicode=true,bookmarksopen=true}

\usepackage{mathtools}
\usepackage{lmodern}
\usepackage[T1]{fontenc}
\usepackage{bbm}
\DeclareMathAlphabet{\mathbfi}{OML}{cmm}{b}{it}
\usepackage{accents}

\let\originalleft\left
\let\originalright\right
\renewcommand{\left}{\mathopen{}\mathclose\bgroup\originalleft}
\renewcommand{\right}{\aftergroup\egroup\originalright}

\makeatletter
\newenvironment{equations}[1][]{\subequations\ifx\relax#1\relax\else\label{#1}\fi\align\ignorespaces}{\endalign\ignorespacesafterend\endsubequations}
\def\@spliteq#1{\begin{equation}\begin{split}#1\end{split}\end{equation}}
\def\splitequation{\collect@body\@spliteq}

\makeatother

\renewcommand{\vec}[1]{{\ifnum9<1#1\mathbf{#1}\else\ifcat\noexpand#1\relax\boldsymbol{#1}\else\mathbfi{#1}\fi\fi}}

\let\oldre\Re
\let\oldim\Im
\renewcommand{\Re}{\oldre\mathfrak{e}\,}
\renewcommand{\Im}{\oldim\mathfrak{m}\,}

\newcommand{\nn}{\nonumber}
\newcommand{\beq}{\begin{equation}}
\newcommand{\eeq}{\end{equation}}

\newcommand{\defeq}{\mathrel{:=}}

\newcommand{\cO}{{\mathcal{O}}}

\usepackage{marginnote}

\newlength{\dhatheight}

\allowdisplaybreaks

\begin{document}

\title{Self-consistency of Conformally Coupled ABJM Theory at the Quantum Level}

\author{Mojtaba Taslimi Tehrani}
\affiliation{Max-Planck-Institut für Mathematik in den Naturwissenschaften, Inselstr. 22, 04103 Leipzig, Germany\\
Institut f\"ur Theoretische Physik, Universit\"at Leipzig,  Br\"uderstr.\ 16, 04103 Leipzig, Germany}

\emailAdd{mojtaba.taslimitehrani@mis.mpg.de}

\abstract{We study the $\mathcal{N}=6$ superconformal Chern-Simons matter field theory (the ABJM theory) conformally coupled to a Lorentzian, curved background spacetime. To support rigid supersymmetry, such backgrounds have to admit twistor spinors.  At the classical level, the symmetry of the theory can be described by a conformal symmetry superalgebra. We show that the full $\mathcal{N}=6$ superconformal algebra persists at the quantum level using the BV-BRST method.
}

\keywords{Quantum field theory in curved space-time, ABJM theory, Conformal symmetry superalgebra, BRST symmetry, Anomaly}

\maketitle

\section{Introduction}

Supersymmetric field theories in curved space-time have attracted attention in recent years \cite{festuccia2011rigid, Jia:2011hw, Samtleben:2012gy, Klare:2012gn, Dumitrescu:2012ha, Cassani:2012ri, Liu:2012bi, Dumitrescu:2012at, Kehagias:2012fh, Hristov:2013spa, Martelli:2012sz}. In such theories, gravity is non-trivial but is kept non-dynamical; it is considered as a rigid background on which matter fields propagate, in contrast with supergravity in which the dynamical gravitational field plays the role of the gauge field of local diffeomorphisms. While theories which preserve rigid supersymmetry in curved space at the classical level are well studied in the literature, the issue of realization of supersymmetry at the quantum level is less investigated. The aim of the present work is to study the ABJM theory \cite{Aharony:2008ug} in Lorentzian curved space-times and to investigate whether this theory can be consistently formulated as a well-defined perturbatively renormalized quantum field theory. In fact, the original theory, formulated in flat space-time, is claimed to remain superconformal at the quantum level. We prove this claim by showing that all the symmetries of this theory are realized at the quantum level, in the sense that we explain in the following.

The ABJM theory \cite{Aharony:2008ug}, usually formulated in three dimensional flat Minkowski space, is an $\mathcal{N}=6$ superconformal gauge theory which plays a crucial role in the $\text{AdS}_4$-$\text{CFT}_3$ correspondence. By conformally coupling the ABJM theory to a curved background $M$, we obtain an action invariant under a \emph{conformal symmetry superalgebra} $\mathcal{S}$ \cite{deMedeiros:2013pps}. $\mathcal{S}$ contains as the even part conformal Killing vectors of $M$ (if any) and $\mathfrak{so}(6)$ (the R-symmetry), and as the odd part the set of all twistor spinors on $M$ in the fundamental representation of $\mathfrak{so}(6)$\footnote{One can also obtain rigid supersymmetric(superconformal) theories by taking the \emph{rigid limit} of off-shell super(conformal)gravity \cite{festuccia2011rigid, Pini:2015xha}.}.

As it turns out, in this theory rigid supersymmetry and local gauge symmetry are intertwined in the sense that supersymmetry transformations close onto field dependent gauge transformations and equations of motion.
The correct gauge-fixing for such ``open algebras'' can be elegantly performed by employing (an extended version of) the BV-BRST formalism \cite{Maggiore:1994dw, Baulieu:2007dk, White:1992wu, Baulieu:2006tu}, which is adapted to curved space-time in \cite{deMedeiros:2013mca}. 
More precisely,  all classical symmetries of the theory, including local gauge symmetry and rigid conformal, R and supersymmetry, can be integrated into one nilpotent BRST symmetry, ${s}$. However, the realization of $s$ at the quantum level is potentially obstructed by anomalies which are subject to certain \emph{consistency conditions} of a cohomological nature. 

We work out the relevant cohomology class which contains potential anomalies and show that this class is trivial in the case of the ABJM theory. This means that the trivial anomaly can be removed by finite renormalization, i.e., by adding suitable finite counter terms to the action. The local and covariant nature of these counter terms in curved space-time is best illustrated in the framework of \emph{locally-covariant QFT} \cite{Brunetti:1995rf, Brunetti:1999jn, Hollands:2001nf, Hollands:2001fb, Hollands:2002ux} which yields a perturbative construction of interacting renormalized QFT on an arbitrary curved (Lorentzian) manifold. 
It is furthermore shown in  \cite{Hollands:2007zg, Tehrani:2017pdx} that the absence of the anomaly directly implies that the full symmetry of the underlying classical theory (in our case, rigid conformal supersymmetry $\mathcal{S}$ and the local gauge symmetry) are preserved at the quantum level, in the sense that there exists a renormalization scheme in which:
(i) the renormalized Noether current of BRST symmetry is conserved, as an operator-valued distribution, (ii) the renormalized BRST charge $Q_L$ is nilpotent and can thus be used to define the Hilbert space of physical state as $\text{Ker} Q_L/ \text{Im} Q_L$,
(ii) the scattering matrix commutes with $Q_L$, and
(iv) for classical observables $\mathcal{O}$ which are annihilated by $s$, the corresponding renormalized composite quantum field $\mathcal{O}_L$ (anti-)commutes with $Q_L$.
\section{ABJM theory conformally coupled to a curved space-time}\label{ABJMCST}

The model written by Aharony, Bergman, Jafferies, and Maldacena (ABJM) \cite{Aharony:2008ug} is a well-studied example of AdS-CFT correspondence. It is a 2+1 dimensional, $\mathcal{N}=6$ superconformal gauge theory with gauge group $G= U(N) \times U(N)$, which consists of two Chern-Simons theories, corresponding to each $U(N)$ factor, with opposite levels (couplings) $k$. 
In this section, we describe the conformally coupled theory. 
\subsection{Conformal symmetry superalgebra $\mathcal{S}$}

The algebraic structure underlying rigid supersymmetric theories is that of a Lie superalgebra $\mathcal{S}$. For the case of the conformally coupled theories, $\mathcal{S}$ is called a \textit{conformal symmetry superalgebra}. This notion was defined in \cite{deMedeiros:2013mca} \cite{deMedeiros:2013pps} where a complete classification of $\mathcal{S}$ for dimensions 3,4,5,6 and for both Riemannian and Lorentzian case was also given. 
Here we describe $\mathcal{S}$ for the case of the 3 dimensional ABJM theory conformally coupled to a curved space-time background. We begin with the geometrical setting.

 Let $(M, g)$ be a Lorentzian 3 dimensional manifold with signature $(-,+,+)$ which admits a spin structure. Given the Levi-Civita connection $\nabla$, the Lie derivative of a vector field $X(x) \in \mathfrak{X}(M)$ along another vector field $Y$ is defined by 
\beq 
 \mathcal{L}_X Y= \nabla_X Y -\nabla_Y X = [X,Y].
\eeq 
It extends to tensor fields of arbitrary rank via the Leibniz rule.  The Lie derivative also defines a representation of $\mathfrak{X}(M)$ on the space of all tensor fields, since $[\mathcal{L}_X, \mathcal{L}_Y] = \mathcal{L}_{[X,Y]}$.
A \textit{conformal Killing vector field} $X$ is defined to satisfy
\begin{equation}
\mathcal{L}_X g + 2 \sigma_X g=0,
\end{equation}
 where $\sigma_X := - \frac{1}{3} \nabla_\mu X^\mu$. That is, the infinitesimal diffeomorphism generated by a conformal Killing vector (CKV) leaves the metric invariant up to a conformal factor $- 2 \sigma_X$.   The linear space of all CKVs together with the Lie bracket of vector fields form the Lie algebra $\mathfrak{X}^c(M)$ of conformal isometries of $(M,g)$.  $\mathfrak{X}^c(M)$ depends only on the conformal class of $g$. For instance, for conformally flat metrics it is isomorphic to $\mathfrak{X}^c(\mathbb{R}^{1,2}) = \mathfrak{so}(2,3)$. 
 
The fermionic counterpart of a CKV is a \emph{conformal Killing spinor} or a \emph{twistor} spinor. In order to introduce them, we first need to fix some basic notions about the spinors. On a curved space-time, each $\gamma_\mu$ is a section of the Clifford bundle $C\ell (TM)$ over $M$, associated to a basis element $\partial_\mu$ in $\mathfrak{X}(M)$. 
 They obey\footnote{For conventions about representations etc. related to spinors in curved space-time see e.g. \cite{deMedeiros:2013pps}}
\begin{equation}
\gamma_\mu \gamma_\nu + \gamma_\nu \gamma_\mu = 2 g_{\mu \nu}\textbf{1}.
\end{equation}

A \textit{twistor spinor} $\epsilon(x)$ is defined by
\begin{equation}\label{twistor}
(\nabla_\mu - \frac{1}{3} \gamma_\mu {\not}\nabla) \epsilon(x)=0,
\end{equation}
where $\nabla_\mu$ is the covariant derivative on spinors.
We denote the linear space of all twistor spinors by $\mathfrak{S}^c(M)$. Similar to $\mathfrak{X}^c(M)$, $\mathfrak{S}^c(M)$ also depends only on the conformal class of $g$, since the twistor equation \eqref{twistor} is Weyl invariant. An important property of twistor spinors is that they ``square to'' conformal Killing vectors in the sense that given any two twistor spinors $\psi, \chi$, the bi-linear $\bar{\psi} \gamma_\mu \chi$ is a conformal Killing vector.

There is a similar notion of Lie derivative acting on spinor fields given by
\begin{equation}
\hat{\mathcal{L}}_X \psi= \mathcal{L}_X \psi + \frac{1}{2} \sigma_X \psi,
\end{equation}
where $\mathcal{L}_X \psi = \nabla_X \psi + \frac{1}{4} (\nabla_\mu X_\nu) \gamma^{\mu \nu}\psi$. 
 $\hat{\mathcal{L}}_X$ defines a representation of the conformal isometry algebra on $\mathfrak{S}^c(M)$, i.e. for $X,Y \in \mathfrak{X}^c(M)$, it satisfies $[\hat{\mathcal{L}}_X, \hat{\mathcal{L}}_Y]= \hat{\mathcal{L}}_{[X,Y]}$.

In the following, we take the supersymmetry parameter, $\epsilon$, to be a real bosonic (commuting) twistor spinor carrying the $\textbf{6}$ representation of $\mathfrak{so}(6)$. 
Let $\{ e_I\}$ be a basis on $\mathbb{C}^6$, with $I =1,2,3,4,5,6$, and let $\{ e^I\}$ be the dual basis. Relative to this basis, $\epsilon= \epsilon^I e_I$. We denote the components of the $\mathfrak{so}(6)$-invariant nondegenerate bilinear form on $\mathbb{C}^6$ by $\epsilon_{IJ}$, with inverse $\epsilon^{IJ}$ (i.e. $\epsilon_{IK} \epsilon^{KJ}= {\delta_I}^J$). 

Now, the conformal symmetry superalgebra, for the conformally coupled ABJM theory, is defined to be a $\mathbb{Z}_2$-graded vector space  $\mathcal{S} = \mathcal{B} \oplus \mathcal{F}$, where
\begin{equation}
\mathcal{B} = \mathfrak{X}^c(M) \oplus \mathfrak{so}(6), \hspace{5 mm} \mathcal{F}_{\mathbb{C}}=  \mathfrak{S}^c(M) \otimes \mathbb{C}^6,
\end{equation}
 together with the following graded Lie bracket for all $X, Y \in \mathfrak{X}^c(M), \rho, \sigma \in   \mathfrak{so}(6), \epsilon_1 , \epsilon_2 \in \mathcal{F}$,
\begin{alignat}{3}\label{gradedLie1}
    [ X \oplus \rho, Y \oplus \sigma ] &=[X,Y]  \oplus [\rho, \sigma], \\ \label{gradedLie2}
    [X \oplus \rho, \epsilon] &= \hat{\mathcal{L}}_X \epsilon+ \rho_\epsilon \cdot \epsilon,  \\ \label{gradedLie3}
    [ \epsilon_1, \epsilon_2]  &=  \xi_{12} \oplus {\rho}_{12}.
\end{alignat}
In the right hand side of \eqref{gradedLie1}, the brackets are the Lie brackets of vector fields and the Lie algebra $\mathfrak{so}(6)$, respectively. In the right hand side of \eqref{gradedLie2}, $(\rho_\epsilon \cdot \epsilon)^I =  {\rho^I}_J \epsilon^J$, with $\rho_\epsilon= {\rho^I}_J e_{I} \otimes e^J$ and 
\beq
\label{rho-IJ}
{\rho^I}_J = \frac{2}{3}  i \bar{\epsilon}^I {\not}\nabla \epsilon_J
\eeq
which is shown to be covariantly constant 
\beq
\label{nabla-rho=0}
\nabla_\mu {\rho^I}_J =0,
\eeq
using the twistor spinor equation \eqref{twistor}.
Also in the right hand side of \eqref{gradedLie3}, $\xi_{12} \equiv \frac{1}{2}(\xi_{\epsilon_1 + \epsilon_2} - \xi_{\epsilon_2} -\xi_{\epsilon_1})$, where  $\xi_\epsilon= \xi^\mu \partial_\mu$ with  
\beq
\label{xi_mu}
\xi^\mu = i \bar{\epsilon}^I \gamma^\mu \epsilon_I \in \mathfrak{X}^c(M),
\eeq
and $\rho_{12} \equiv \frac{1}{2}(\rho_{\epsilon_1 + \epsilon_2} - \rho_{\epsilon_2} -\rho_{\epsilon_1})$ which is constant on $M$, as required by consistency of \eqref{gradedLie3}. The graded Lie bracket $[-,-]$ can be shown to satisfy the graded Jacobi identity \cite{deMedeiros:2013pps}.

\subsection{Rigid backgrounds admitting twistor spinors}

Admitting twistor spinors puts a strong restriction on the underlying background space-time. A complete classification of such Lorentzian manifolds is given in \cite{baum2004twistor, Baum:2002jt, baum2012holonomy, leitner2005conformal, baum2008conformal}. The maximum number of linearly independent Killing vectors, and twistor spinors in $d$ dimensions are  $\frac{1}{2}(d+1)(d+2)$ and $2^{\lfloor \frac{d}{2}\rfloor+1 }$ respectively. The two bounds are saturated for locally conformally flat metrics.
In three dimensions, Lorentzian manifolds admitting twistor spinors fall into two distinct classes:
\begin{itemize}
\item[(1)] Locally conformally flat metrics\footnote{In three dimensions, the Weyl tensor vanishes identically. However, the Cotton-York tensor
$C_{\mu \nu \rho} = \nabla_\mu K_{\nu \rho} - \nabla_\nu K_{\mu \rho}$ is in general non-zero, and vanishes if and only if the metric is conformally flat.}, such as $\mathbb{R}^{1,2}, dS_3$ and $AdS_3$. On Minkowski $\mathbb{R}^{1,2}$, the general solution to the twistor equation is 
\beq
\label{twistor-Minkowski}
\epsilon = \alpha \epsilon_0 + \beta {\not}x \epsilon_0,
\eeq
with $\epsilon_0$ begin a constant spinor with 4 linearly independent components, and $\alpha, \beta$ arbitrary constants. There are also $10$ linearly independent conformal Killing vectors which form the algebra $\mathfrak{so}(2,3)$. The ABJM superalgebra is $\mathcal{S}_{\mathbb{R}^{1,2}} \cong \mathfrak{osp}(6|4)$.

\item[(2)] Certain types of pp-wave metrics \cite{Jordan2009, baum2004twistor} which are not conformally flat.  The pp-wave metric in Brinkmann coordinates $(u,v,x)$ \cite{Brinkmann1925} takes the form
\begin{equation}
g = 2 du dv + h(u, x) du^2 + dx^2,
\end{equation}
where $h$ is an arbitrary smooth function of $u$ and $x$. All pp-wave metrics whose $h$ are higher than quadratic in $x$ are not conformally flat. 

For pp-waves with arbitrary $h$, any twistor spinor $\epsilon$ is \emph{parallel}, i.e. $\nabla_\mu \epsilon=0$, and for the non-conformally flat class, the maximum number of independent components are 2. The only conformal Killing vector is $\xi=\partial_\nu$ which is null.

\end{itemize} 
\subsection{Supersymmetry transformations and the supersymmetric Lagrangian}\label{SUSYtrafo}
Besides rigid conformal supersymmetry, the ABJM theory has a local gauge symmetry with gauge group $G= U(N)\times U(N)$, with Lie algebra denoted by $\mathfrak{g}=\mathfrak{u}(N)\oplus \mathfrak{u}(N)$. Here we construct an invariant Lagrangian under $\mathcal{S} \oplus \mathcal{G}$, where $\mathcal{G}= C^\infty(M,\mathfrak{g})$ in component form. It differs from the flat space Lagrangian \cite{Aharony:2008ug, Bandres:2008ry} by expected curvature couplings. 

\subsubsection*{Field content}
The basic fields are $(A, \hat{A}, \mathbb{X}, \Psi)$ as explained in the following.\footnote{We closely follow the notation of \cite{Bandres:2008ry}.} 

\begin{itemize}
\item $(A, \hat{A})= (A_\mu dx^\mu, \hat{A}_\mu dx^\mu)$, the gauge connection, is a $\mathfrak{u}(N)\oplus \mathfrak{u}(N)$-valued 1-form. 

\item Matter fields, $(\mathbb{X}_A, \Psi_A)$, with $A=1,2,3,4$, are a complex scalar and a Dirac spinor field respectively. They both carry the \textbf{4} representation under $\mathfrak{su}(4)\cong \mathfrak{so}(6)$ R-symmetry. 
$\mathbb{X}_A$ and $\Psi_A$ also transform in the bi-fundamental representation $( \bar{\textbf{N}}, \textbf{N})$ of  $\mathfrak{u}(N)$, while $\mathbb{X}^A$ and $\Psi^A$ transform in the anti-bi-fundamental representation $( \textbf{N}, \bar{\textbf{N}})$. 
\end{itemize}

Since matter fields $\mathbb{X}^A, \Psi^A$ and the supersymmetry parameters $\epsilon^I$ carry different representations of the R-symmetry, one needs intertwiners between the $\textbf{6}$ and $\textbf{4}$ representations. These are six $4 \times 4$ anti-symmetric matrices $\Gamma^I_{AB} = - \Gamma^I_{BA}$ satisfying
\begin{equation}
\Gamma^I \tilde{\Gamma}^J + \Gamma^J \tilde{\Gamma}^I = 2 \delta^{IJ} \textbf{1},
\end{equation}
where $\tilde{\Gamma}^{IAB} = \frac{1}{2} \epsilon^{ABCD}\Gamma^I_{CD}$, where $\epsilon^{ABCD}$ is the $\mathfrak{so}(6)$-invariant totally anti-symmetric tensor in the \textbf{4} representation. We also denote $({\Gamma^I}_J)_A^C \equiv  \epsilon_{JK}\Gamma^{[I}_{AB} \tilde{\Gamma}^{K]BC}$, and we find  $\Gamma^{(I}_{AB} \tilde{\Gamma}^{J)CD} = -\frac{1}{3}\delta_{AB}^{CD} \delta^{IJ}$.
\subsubsection*{Action of $\mathcal{S}$ on fields}
For the \textbf{bosonic part $\mathcal{B} \subset \mathcal{S}$}, the action of a conformal Killing vector $X$ on the space of field configurations is given by 
\beq
\label{delta_X}
\delta_X \Phi = \mathcal{L}_X \Phi + w_\Phi \sigma_X \Phi,
\eeq 
where $w_\Phi$ is the Weyl weight of a generic field $\Phi$ given in Table \ref{table1}. 
\begin{table}
\begin{center}
    \begin{tabular}{   | l | l | l | l | l | l | p{0.4 cm}  | }
    \hline
    \textit{Fields} & $(A_\mu , \hat{A}_\mu)$ & $\mathbb{X}_A$ &$\Psi_A$  \\ \hline \hline 
    Dimension & $1$  &   $1/2$& $1$ \\ \hline
   Weyl weight & $0$ &   $-1/2$ & $-1$ \\ \hline
   Spin & $1$  &   $1$ & $1/2$ \\ \hline
    \end{tabular}
    \caption{\label{table1} Basic fields and their data.}
\end{center}
\end{table}

In addition, the $\mathfrak{so}(6)$ R-symmetry acts by 
\begin{align}
\label{delta-rho}
\delta_\rho A_\mu =0 = \delta_\rho \hat{A}_\mu, \quad \delta_\rho \mathbb{X}_A = {\rho^B}_A \mathbb{X}_B, \quad \delta_\rho \Psi_A = {\rho^B}_A \Psi_B,
\end{align}
where ${\rho^B}_A = ({\Gamma^I}_J)_A^B  {\rho^J}_I$, and $\delta_\rho$ satisfies $[\delta_\rho, \delta_{\sigma}]= \delta_{[\rho, \sigma]}$.
Since the ${\rho^I}_J$ are covariantly constant \eqref{nabla-rho=0}, the variations \eqref{delta_X} and \eqref{delta-rho} define a representation of $\mathcal{B}$ on the space of fields since
\begin{equation}
[\delta_X + \delta_{\rho}, \delta_Y + \delta_{\sigma}] = \mathcal{L}_{[X,Y]} +w_\Phi \sigma_{[X,Y]}+ \delta_{[\rho, \sigma]}= \delta_{[X, Y]} + \delta_{[\rho, \sigma]} . 
\end{equation}
 The action of the \textbf{Fermionic part $\mathcal{F} \subset \mathcal{S}$} is the curved space ganeralization of the usual  superconformal transformations of the ABJM theory in flat spacetime \cite{Bandres:2008ry}:
\begin{equations}[SUSY-variations]
 & \delta_\epsilon A_\mu  =   \Gamma^I_{AB}\bar{\epsilon}^I \gamma_\mu \Psi^A \mathbb{X}^B - \tilde{\Gamma}^{IAB} \mathbb{X}_B \bar{\Psi}_A \gamma_\mu \epsilon^I, \\ 
&\delta_\epsilon \hat{A}_\mu =\Gamma^I_{AB}\mathbb{X}^B \bar{\epsilon}^I \gamma_\mu \Psi^A  - \tilde{\Gamma}^{IAB}\bar{\Psi}_A \gamma_\mu \epsilon^I \mathbb{X}_B, \\  
& \delta_\epsilon \mathbb{X}_A  =   i \Gamma^I_{AB} \bar{\epsilon}^I\Psi^B,  \\
& \delta_\epsilon \Psi_A = \Gamma^I_{AB} {\not}D \mathbb{X}^B \epsilon^I   + {\Gamma}^I_{AB} (\mathbb{X}^C \mathbb{X}_C \mathbb{X}^B - \mathbb{X}^B \mathbb{X}_C \mathbb{X}^C) \epsilon^I - 2 {\Gamma}^I_{BC} \mathbb{X}^B \mathbb{X}_A \mathbb{X}^C \epsilon^I  - {\frac{1}{3} \Gamma^I_{AB} {\not}\nabla \epsilon^I \mathbb{X}^B},   
\end{equations}
where $\epsilon^I$ is a twistor spinor satisfying \eqref{twistor}. The superconformal variations \eqref{SUSY-variations} define a representation of $\mathcal{F}$ on fields only up to a field-dependent gauge transformation and equations of motion, by contrast to the case of $\mathcal{B}$. In fact, the commutator of two supersymmetry transformations turns out to be of the form
\footnote{\label{footnote}For our commuting supersymmetry parameters $\epsilon_1$, $\epsilon_2$, the (anti-) commutator $[\delta_{\epsilon_1}, \delta_{\epsilon_2}] = \delta_{\epsilon_1}\delta_{\epsilon_2} + \delta_{\epsilon_2} \delta_{\epsilon_1}$ is related to $\delta^2_{\epsilon}$ via polarization, i.e. $[\delta_{\epsilon_1}, \delta_{\epsilon_2}]= \delta^2_{\epsilon_1 + \epsilon_2} - \delta^2_{\epsilon_1}- \delta^2_{\epsilon_2}$. In verifying \eqref{deltaepsilon2}, we have used $\bar{\epsilon} \epsilon =0$ and $\epsilon \bar{\epsilon} = \frac{1}{2} {\not} \xi$ as well as
${\not}{\nabla}^2 \epsilon  = -\frac{3}{8} R \epsilon $
which follow from the twistor equation for $\epsilon$, together with the Lichnerowicz-Weitzenb\"{o}ck identity ${\not}{\nabla}^2 \psi =  \nabla^2 \psi + \frac{1}{4} R \psi$ for arbitrary spinors $\psi$.}

\begin{equation}\label{deltaepsilon2}
\delta_\epsilon^2 =  \delta_{\xi(\epsilon)} +  \delta_{\rho(\epsilon)} + \delta_{(\Lambda(\phi), \hat{\Lambda}(\phi))} + \delta_{\text{e.o.m.}} ,
\end{equation}
where $\delta_{\xi(\epsilon)} = \mathcal{L}_\xi + w_\Phi \sigma_\xi$ with $\xi^\mu$ defined in \eqref{xi_mu}, where $\rho(\epsilon)$ is defined in \eqref{rho-IJ}, and where $\delta_\rho$ is defined in \eqref{delta-rho}. The field-dependent gauge transformation parameters $(\Lambda, \hat{\Lambda}) \in C^\infty(\mathcal{C}) \otimes(\mathfrak{u}(N) \oplus \mathfrak{u}(N))$ in this equation are
\begin{align}
{\Lambda}(\phi)& =  (\xi^\mu A_\mu)  - (\bar{\epsilon}_I({\Gamma^I}_J)_D^C\epsilon^J \mathbb{X}^D \mathbb{X}_C) \\
{\hat{\Lambda}}(\phi) &=  (\xi^\mu \hat{A}_\mu)  - (\bar{\epsilon}_I({\Gamma^I}_J)_D^C\epsilon^J  \mathbb{X}_C \mathbb{X}^D ) ,
\end{align}
and $\delta_{(\Lambda, \hat{\Lambda})}$ is defined in equation \eqref{gaugetrafo} below. The $\delta_{\text{e.o.m.}}$-term is non-trivial only for the the fermionic field $\Psi_A$ and the gauge fields $A_\mu$, $\hat{A}_\mu$. It is given by
\begin{align}
 \delta_{\text{e.o.m.}} A_\mu &= \xi_\mu (F_{\mu \nu} - \epsilon_{\mu\nu \rho} J^\rho),\\
 \delta_{\text{e.o.m.}} \hat{A}_\mu &= \xi_\mu (\hat{F}_{\mu \nu} - \epsilon_{\mu\nu \rho} \hat{J}^\rho),\\
 \delta_{\text{e.o.m.}} \Psi_A &= {\not}\xi E_A +  ({\Gamma^I}_J)^B_A \bar{\epsilon}_I \epsilon^J E_B,
\end{align}
where 
\begin{align}
J^\mu &= i \mathbb{X}_A D^\mu \mathbb{X}^A - i (D^\mu \mathbb{X}_A)\mathbb{X}^A - \bar{\Psi}^A \gamma^\mu \Psi_A,\\
\hat{J}^\mu &= i \mathbb{X}^A D^\mu \mathbb{X}_A - i (D^\mu \mathbb{X}^A)\mathbb{X}_A - \bar{\Psi}_A \gamma^\mu \Psi^A, \\
E_A &= {\not} D \Psi_A - 2 \epsilon_{ABCD} \mathbb{X}^B \Psi^C \mathbb{X}^D - \mathbb{X}^B\mathbb{X}_B \Psi_A + \Psi_A \mathbb{X}_B \mathbb{X}^B  - 2 \Psi_B \mathbb{X}_A \mathbb{X}^B + 2 \mathbb{X}^B \mathbb{X}_A \Psi_B.
\end{align}

In the particular case of the Minkowski spacetime $\mathbb{R}^{1,2}$, the solution to the twistor spinor equation \eqref{twistor-Minkowski} is a linear combination of a constant spinor $\epsilon_0^I$, and ${\not} x \epsilon_0^I \equiv \eta^I$. In this case, the above transformations for $\epsilon^I= \epsilon^I_0$ reduce to the ABJM  supersymmetry transformations on $\mathbb{R}^{1,2}$, while with $\epsilon^I= \eta^I$, they reduce to the superconformal transformations, both derived in \cite{Bandres:2008ry}.

\subsubsection*{Superconformal Lagrangian}
To obtain an invariant Lagrangian, we also need to specify how the dynamical fields transform under local gauge symmetry.  
Under a local gauge transformation with parameters $(\Lambda, \hat{\Lambda}) \in C^\infty(M) \otimes(\mathfrak{u}(N) \oplus \mathfrak{u}(N))$,
\begin{align} \label{gaugetrafo}
\delta_\Lambda A_\mu &= D_\mu \Lambda = \partial_\mu \Lambda + i [A_\mu, \Lambda],\\
\delta_{\hat{\Lambda}} \hat{A}_\mu &= D_\mu \hat{\Lambda} = \partial_\mu \hat{\Lambda} + i [\hat{A}_\mu, \hat{\Lambda}],\\
\delta_{\Lambda \hat{\Lambda}}\mathbb{X}_A &= -i \Lambda \mathbb{X}_A + i \mathbb{X}_A \hat{\Lambda},\\
\delta_{\Lambda \hat{\Lambda}}\Psi^A &= -i \Lambda \Psi^A + i \Psi^A \hat{\Lambda}.
\end{align}
The gauge covariant derivatives act on fields by
\begin{align}
D_\mu \mathbb{X}_A &= \nabla_\mu \mathbb{X}_A + i (A_\mu \mathbb{X}_A - \mathbb{X}_A \hat{A}_\mu),\\
D_\mu \mathbb{X}^A &= \nabla_\mu \mathbb{X}^A + i (\hat{A}_\mu \mathbb{X}^A - \mathbb{X}_A A_\mu).
\end{align}
The invariant Lagrangian, $\mathcal{L}_{\mathcal{N}=6}$, under the action of $\mathcal{S} \oplus \mathcal{G}$ is the conformally coupled version of the flat space Lagrangian given in \cite{Aharony:2008ug, Bandres:2008ry}. Explicitly,
\begin{equation}
\mathcal{L}_{\mathcal{N}=6}= \frac{k}{2 \pi} \big( \mathcal{L}_{\text{CS}}+ \mathcal{L}_{\text{kin}}+  \mathcal{L}_{\text{int}} \big) , 
\end{equation}
with $k$ being the Chern-Simons level, where
\begin{align}
\mathcal{L}_{\text{CS}}&= \epsilon^{\mu \nu \rho} \text{Tr} \left( \frac{1}{2}A_\mu \nabla_\nu A_\rho - \frac{1}{2}\hat{A}_\mu \nabla_\nu \hat{A}_\rho + \frac{i}{3} A_\mu A_\nu A_\rho -  \frac{i}{3} \hat{A}_\mu \hat{A}_\nu \hat{A}_\rho   \right),\\ 
\mathcal{L}_{\text{kin}}&=  \text{Tr} \left( -  {{D}_\mu} \mathbb{X}^A  {{D}^\mu} \mathbb{X}_A + i \bar{\Psi}_A {\not} {{D}} \Psi^A  -\frac{1}{8} R \mathbb{X}^A \mathbb{X}_A \right), \\
\mathcal{L}_{\text{int}} &= \text{Tr} \left( \frac{i}{2} (\bar{\Psi}^{2 A}\mathbb{X}^{2}_{A}   + \bar{\Psi}_A^{2}\mathbb{X}^{2 A} ) +  \frac{1}{12} (\mathbb{X}_A^{3 I} \mathbb{X}_I^{3A} + \mathbb{X}_I^{3A} \mathbb{X}_A^{3 I} )\right),
\end{align}
where $R$ is the scalar curvature of $(M,g)$, and where 
\begin{align} \label{Psi2Phi2}
&\bar{\Psi}^{2A} \mathbb{X}_A^2 \equiv   -  \epsilon_{ABCD}(\bar{\Psi}^A \mathbb{X}^B \Psi^C \mathbb{X}^D)+ (\bar{\Psi}^A \Psi_A \mathbb{X}_B \mathbb{X}^B) - 2 (\bar{\Psi}^A \Psi_B \mathbb{X}_A \mathbb{X}^B), \\
&\bar{\Psi}_A^2 \mathbb{X}^{2A} \equiv   \epsilon^{ABCD}(\bar{\Psi}_A \mathbb{X}_B \Psi_C \mathbb{X}_D))-   (\bar{\Psi}_A \Psi^A \mathbb{X}^B \mathbb{X}_B) +  2(\bar{\Psi}_A \Psi^B \mathbb{X}^A \mathbb{X}_B),\\
& \mathbb{X}^{3IA} \equiv \tilde{\Gamma}^{IAB} ( \mathbb{X}_C \mathbb{X}^C \mathbb{X}_B  -  \mathbb{X}_B \mathbb{X}^C \mathbb{X}_C) - 2 \tilde{\Gamma}^{IBC} \mathbb{X}_B \mathbb{X}^A \mathbb{X}_C,\\ \label{Phi3}
& \mathbb{X}^{3I}_A \equiv   \Gamma^I_{AB} (\mathbb{X}^C \mathbb{X}_C \mathbb{X}^B - \mathbb{X}^B \mathbb{X}_C \mathbb{X}^C) - 2 \Gamma_{BC}^I \mathbb{X}^B \mathbb{X}_A \mathbb{X}^C.
\end{align}
The Lagrangian $\mathcal{L}_{\mathcal{N}=6}$ is manifestly $\mathfrak{so}(6)$ invariant. The action is gauge, Weyl and supersymmetry invariance of the action can be checked using \eqref{twistor}, the identities in footnote \ref{footnote} and algebraic spinor identities such as $\bar{\psi} \chi = \bar{\chi} \psi$ and $\bar{\psi} \gamma^\mu \chi = - \bar{\chi} \gamma^\mu \psi$
for anti-commuting $\psi, \chi$, as well as the Fierz identity $\chi \bar{\psi}  = -\frac{1}{2} ( \bar{\psi} \chi \textbf{1} + (\bar{\psi} \gamma^\mu \chi) \gamma_\mu)$.
\section{Gauge fixing and BRST symmetry}\label{Gauge-fixing-BRST}
In this section, we describe the gauge-fixing of the the conformally coupled ABJM theory. As can be seen from \eqref{deltaepsilon2}, in this theory, supersymmetry transformations not only close onto local gauge transformations but also onto a term proportional to the equations of motion of fields. This defines a so-called ``open algebra'' whose gauge-fixing requires a suitable extended version of the BV-BRST formalism \cite{Maggiore:1994dw, Baulieu:2007dk, White:1992wu, Baulieu:2006tu, deMedeiros:2013mca}. Here, we work this out for the theory at hand.

One associates besides the dynamical ghost for the local gauge symmetry, a set of rigid ghosts for conformal, R, and supersymmetry as well. Moreover, to every field and ghost $\Phi$ of the theory, there is an associated anti-field $\Phi^\ddag$. Then an extended BRST differential ${s}$ acting on $(\Phi, \Phi^\ddag)$ is defined roughly as follows. $ {s} \Phi$ is the sum of all (local and rigid) symmetry transformations of $\Phi$ with each symmetry parameter replaced by its corresponding ghost, plus certain terms needed to incorporate the open nature of supersymmetry algebra. ${s} \Phi^\ddag$ is the equation of motion of $\Phi$ coming from an extended action ${S}$. Of course, ${S}$ and the precise form of ${s}$ has to be made in such a way to satisfy ${s}^2=0$ and ${s} {S}=0$ which we intend to explain in the following.
\subsection{BRST structure of $\mathcal{N}=6$ superconformal Chern-Simons matter theory}\label{enlargedtheory}
 
In order to perform the gauge fixing and the BRST quantization, we need to enlarge the field configuration space, containing $(A_\mu, \hat{A}_\mu, \mathbb{X}_A, \Psi_A)$, to include the following elements:
\begin{itemize}
\item A pair $(c, \hat{c})$ of $(\mathfrak{u}(N) \oplus \mathfrak{u}(N))$-valued Grassmann odd scalars, which are dynamical ghosts for the local gauge symmetry parameters. In components, $c= c^i T_i$, $\hat{c}= \hat{c}^i \hat{T}_i$ (see Section \ref{SUSYtrafo}).

\item Non-dynamical ghosts ($X_\mu, {\alpha^I}_J, \epsilon^I$), associated to the rigid symmetries; conformal, R, and supersymmetry.

\item A $(B, \hat{B})$ and $(\bar{c}, \bar{\hat{c}})$ system which are two pairs of $(\mathfrak{u}(N) \oplus \mathfrak{u}(N))$-valued dynamical scalar fields needed for the gauge fixing term.

\item Anti-fields $\Phi^\ddag$ for each field and ghost $\Phi$. For instance $A_\mu^\ddag$ for $A_\mu$ etc.
By definition, each ${\Phi}^\ddag$ has the opposite Grassmann parity of the $\Phi$, and if ($\Delta_\Phi, g_\Phi, w_\Phi$) are dimension, ghost number and Weyl weights of $\Phi$ respectively, then $\Phi^\ddag$ has 
\beq
(\Delta_{\Phi^\ddag}, g_{\Phi^\ddag}, w_{\Phi^\ddag})= (3- \Delta_\Phi, -1- g_\Phi, -3- w_\Phi).
\eeq

\end{itemize}

We summarize the data for all ghosts in Table \ref{table2}.
\begin{table}[h]
\begin{center}
    \begin{tabular}{  | l | l | l | l | l | l | l  | l | l | l | l | l | p{0.4 cm}  | } 
    \hline
    \textit{Ghosts} & $(c, \hat{c})$ &$\alpha^{IJ}$ &$\epsilon^{I }$  & $X_\mu$ & $(\bar{c}, \bar{\hat{c}})$  & $(B, \hat{B})$ \\ \hline \hline 
    Dimension & $0$ &  $0$& $- 1/2$& $-1$ & $1$& $1$    \\ \hline
    Ghost number & $1$  &  $1$ & $1$ & $1$ & $-1$& $0$   \\ \hline
   Grassmann parity & $1$  &  $1$ & $0$ & $1$ &  $1$& $0$   \\  \hline
   Weyl weight & $0$  &  $0$ & $1/2$&  $0$ &  $-1$ &  $-1$     \\ \hline
   Dynamical & Yes  &No &No & No & Yes & Yes   \\ \hline
    \end{tabular}
    \caption{\label{table2}Data for ghosts and gauge-fixing fields.}
\end{center}
\end{table}
Let us first define the nilpotent operator $\hat{s}$ which increases the ghost number by $+1$, and acts on all fields $\Phi$ by
\begin{equations}[sAmu]
&\hat{s}  A_\mu = \mathcal{D}_\mu c + \mathcal{L}_X A_\mu + \Gamma^I_{AB}\bar{\epsilon}^I \gamma_\mu \Psi^A \mathbb{X}^B - \tilde{\Gamma}^{IAB} \mathbb{X}_B \bar{\Psi}_A \gamma_\mu \epsilon^I  + \epsilon_{\mu \nu \rho} \xi^\nu A^{\ddag \rho} , \\ 
&\hat{s}  \hat{A}_\mu = \mathcal{D}_\mu \hat{c} + \mathcal{L}_X \hat{A}_\mu + \Gamma^I_{AB}\mathbb{X}^B \bar{\epsilon}^I \gamma_\mu \Psi^A  - \tilde{\Gamma}^{IAB}\bar{\Psi}_A \gamma_\mu \epsilon^I \mathbb{X}_B +  \epsilon_{\mu \nu \rho} \xi^\nu \hat{A}^{\ddag \rho} ,\\
&\hat{s} \mathbb{X}_A = -i c\mathbb{X}_A + i \mathbb{X}_A \hat{c}  + ({\Gamma^I}_J)_A^B{\alpha^J}_I \mathbb{X}_B  +  ( \mathcal{L}_X {-\frac{1}{2}} \sigma_X) \mathbb{X}_A +  i \Gamma^I_{AB} \bar{\epsilon}^I\Psi^B, \\ \nonumber \label{s Psi}
&\hat{s}  \Psi_A  =   -i c\Psi_A + i \Psi_A \hat{c}  + ({\Gamma^I}_J)_A^B {\alpha^J}_I \Psi_B +  ( \mathcal{L}_X { -}\sigma_X) \Psi_A - {\frac{1}{3} \Gamma^I_{AB} {\not}\nabla \epsilon^I \mathbb{X}^B}  + \Gamma^I_{AB} {\not}D \mathbb{X}^B \epsilon^I \\
& \quad  \quad + {\Gamma}^I_{AB} (\mathbb{X}^C \mathbb{X}_C \mathbb{X}^B - \mathbb{X}^B \mathbb{X}_C \mathbb{X}^C) \epsilon^I - 2 {\Gamma}^I_{BC} \mathbb{X}^B \mathbb{X}_A \mathbb{X}^C \epsilon^I - {\not} \xi \bar{\Psi}_A^{\ddag} - ({\Gamma^I}_J)^B_A \bar{\epsilon}_I \epsilon^J \bar{\Psi}^\ddag_B, \\
&\hat{s} c =  - \frac{1}{2} [c,c] -   \mathcal{L}_X c -  {\frac{1}{2}((\xi^\mu A_\mu)  - (\bar{\epsilon}_I({\Gamma^I}_J)_D^C\epsilon^J \mathbb{X}^D \mathbb{X}_C) ) },\\
&\hat{s} \hat{c} =  - \frac{1}{2} [\hat{c}, \hat{c}] -   \mathcal{L}_X \hat{c} - {\frac{1}{2} ((\xi^\mu \hat{A}_\mu)  - (\bar{\epsilon}_I({\Gamma^I}_J)_D^C\epsilon^J  \mathbb{X}_C \mathbb{X}^D )) },\\
&\hat{s} {\alpha^I}_J = - {\alpha^I}_K {\alpha^K}_J  + \frac{2}{3}(i\bar{\epsilon}^I {\not}\nabla\epsilon_J),  \\
&\hat{s} X_\mu =  - \frac{1}{2} [X,X]_\mu - \xi_\mu, \\
&\hat{s}  \epsilon^I =  - ( \mathcal{L}_X + \frac{1}{2}\sigma_X) \epsilon^I +  {\alpha^I}_J \epsilon^J , \\ 
&\hat{s} \bar{c} =  ( \mathcal{L}_X - \sigma_X) \bar{c} + B,\\
&\hat{s} \bar{\hat{c}} =  ( \mathcal{L}_X - \sigma_X) \bar{\hat{c}} + \hat{B},\\
&\hat{s}  B =  ( \mathcal{L}_X -  \sigma_X) B +  ( \mathcal{L}_\xi - \sigma_\xi) \bar{c} ,\\
&\hat{s}  \hat{B} =  ( \mathcal{L}_X -  \sigma_X) \hat{B} +  ( \mathcal{L}_\xi -  \sigma_\xi) \bar{\hat{c}}.
\end{equations}
We next define the following action functional $\hat{S}$ 
\begin{equation}\label{hatS}
\hat{S} = S_{\mathcal{N}=6} + S_{\text{sc}} +S_{\text{af}} ,
\end{equation}
where $S_{\mathcal{N}=6} = \int_M \mathcal{L}_{\mathcal{N}=6}$, and
\begin{equation}
S_{\text{sc}} = \frac{k}{2 \pi} \sum_i \int_M \hat{s} \Phi^i \cdot \Phi_i^\ddag.
\end{equation}
is a source term, which couples all BRST transformed fields $\hat{s} \Phi^i$ to their corresponding sources $\Phi_i^\ddag$. 
In addition, we have added the anti-field action $S_{\text{af}}$ which is quadratic in $\Psi^\ddag, A^\ddag,\hat{A}^\ddag$
\begin{equation}
S_{\text{af}} = \frac{k}{4 \pi} \int_M \left(  \Psi^{A \ddag} {\not} \xi  \bar{\Psi}_A^{ \ddag}+ ({\Gamma^I}_J)^B_A \bar{\epsilon}_I \epsilon^J \Psi^{\ddag A}\bar{\Psi}^\ddag_B+  \epsilon_{\mu \nu \rho} {A}^{\ddag \mu }  \xi^\nu {A}^{ \ddag \rho}  + \epsilon_{\mu \nu \rho} \hat{A}^{\ddag \mu }  \xi^\nu \hat{A}^{\ddag \rho } \right).
\end{equation}
The operator $\hat{s}$, can now be extended to anti-fields by $\hat{s} = \llbracket \hat{S},- \rrbracket$, where
\begin{equation}\label{antibracket}
\llbracket \mathcal{O}_1, \mathcal{O}_2 \rrbracket := \frac{\delta \mathcal{O}_1}{\delta \Phi}  \frac{\delta \mathcal{O}_2}{\delta \Phi^\ddag} -  \frac{\delta \mathcal{O}_1}{\delta \Phi^\ddag} \frac{\delta \mathcal{O}_2}{\delta \Phi}.
\end{equation}
is called the \emph{anti-bracket} and satisfies the graded Jacobi identity.

Finally, gauge-fixing is done by performing a ``canonical transformation''
\begin{equation}
\label{eq:canonical-trnsf}
e^{\llbracket-,\Psi \rrbracket} =  \textbf{1} +  \llbracket-,\Psi \rrbracket + \frac{1}{2!} \big \llbracket \llbracket-,\Psi \rrbracket, \Psi \big\rrbracket + \frac{1}{3!} \Big\llbracket \big\llbracket \llbracket-,\Psi \rrbracket, \Psi \big\rrbracket, \Psi \Big\rrbracket + \dots,
\end{equation}
on the action $\hat{S}$, where $\Psi$ is the ``gauge fermion'' with ghost number $-1$ given by 
\begin{equation}\label{gauge fermion}
\Psi =  \frac{k}{2 \pi} \int_M  (\nabla_\mu A^\mu -\frac{1}{2} B) \bar{c} - (\nabla_\mu \hat{A}^\mu -\frac{1}{2} \hat{B}) \bar{\hat{c}}, 
\end{equation}
which implements the Feynman gauge. The gauge-fixed action of the enlarged theory is thus defined to be 
\beq 
\label{eq:gauge-fixed-action}
 S = e^{\llbracket-,\Psi \rrbracket} \hat{S} = \hat{S} + \hat{s} \Psi + \frac{1}{2} \big\llbracket \hat{s} \Psi, \Psi \big\rrbracket.
\eeq
Note that by contrast to the usual gauge-fixing of the Yang-Mills theory, our gauge-fixed action has an additional term $\frac{1}{2}  \big\llbracket \hat{s} \Psi, \Psi \big\rrbracket$. This term appears due to the presence of the quadratic action $S_{\text{af}}$ in anti-fields $\Psi^\ddag_A$ and $(A^\ddag_\mu, \hat{A}^\ddag_\mu)$ which, in turn, is needed for obtaining a nilpotent differential because of the  $\delta_{\text{e.o.m.}}$-term in \eqref{deltaepsilon2} which occurs for $\Psi_A$ and $(A_\mu, \hat{A}_\mu)$. Such a term is characteristic for supersymmetric gauge theories where the algebra of transformations only closes on-shell.

The gauge-fixed action $S$ leads to the BV-BRST differential $s$, defined by 
\beq
\label{s=(S,-)}
{s} \defeq e^{\llbracket-,\Psi \rrbracket} \circ \hat{s} \circ e^{-\llbracket-,\Psi \rrbracket} =  \llbracket S,- \rrbracket.
\eeq
The action of $s$ on all fields coincides with that of $\hat{s}$ \eqref{sAmu}, except for $(A_\mu, \hat{A}_\mu)$ where it is
\begin{align}
s A_\mu = \hat{s} A_\mu +  \epsilon_{\mu \nu \rho} \xi^\nu \nabla^\rho \bar{c}, \qquad s \hat{A}_\mu = \hat{s} \hat{A}_\mu +  \epsilon_{\mu \nu \rho} \xi^\nu \nabla^\rho \hat{\bar{c}}.
\end{align}
On anti-fields is given by 
\begin{equation}
{s} \Phi^\ddag =\llbracket S, \Phi^\ddag \rrbracket = \frac{\delta {S}}{\delta \Phi}.
\end{equation}
The $s$ transformation of those anti-fields which are needed later are given in Appendix \ref{BRSTantifields}. The gauge-fixed action $S$ is checked to satisfy
\beq
\label{master-eq}
\llbracket S,S \rrbracket=0,
\eeq 
and using the graded Jacobi identity for anti-bracket, one shows from this that
\beq
s^2=0,
\eeq
as an operator. The identity \eqref{master-eq} expresses the invariance of $S$ under \emph{all} rigid and local symmetries, i.e. $sS=0$. In fact, the Noether current of $s$ is given by
\beq
\label{BRST-current}
J_\mu = J_\mu^{\text{gauge}} + J_\mu^{\text{SUSY}} + J_\mu^{\text{R}} + J_\mu^{\text{conf}} + J_\mu^{\text{anti-fields}},
\eeq
which includes 
the gauge current, supercurrent, R-symmetry current and conformal current:
\begin{align}
J_\mu^{\text{gauge}} &= \text{Tr}\Big( \frac{1}{2} \epsilon_{\mu \nu \rho}( F^{\nu \rho} c - \hat{F}^{\nu \rho} \hat{c}) \Big), \\
J_\mu^{\text{SUSY}} &=  \text{Tr}\Big(  \delta_\epsilon \bar{\Psi}_A \gamma_\mu \Psi^A +  \delta_\epsilon \bar{\Psi}^A \gamma_\mu \Psi_A   \Big), \\ \nn
J_\mu^{\text{R}} & =  \text{Tr} \Big( D_\mu \mathbb{X}^A   {\rho^B}_A \mathbb{X}_B + D_\mu \mathbb{X}_A  {\rho^A}_B \mathbb{X}^B +  i (\bar{\Psi}_A \gamma_\mu   {\rho^A}_B \Psi^B+ \bar{\Psi}^A \gamma_\mu   {\rho^B}_A \Psi_B ) \Big), \\
J_\mu^{\text{conf}} & = X^\nu T_{\mu \nu}, 
\end{align}
where $T_{\mu \nu}$ is the energy-momentum tensor of $S_{\mathcal{N}=6}$ and  $ J_\mu^{\text{af}}$ is an anti-field dependent part.
\section{Calculation of BRST cohomology}\label{BRST-calculation}

In this part, we present the cohomology analysis of the nilpotent BRST differential $s$ defined in the previous section. We calculate a particular cohomology class which contains the potential anomalies of the classical symmetry at quantum level, and show that this class is trivial in our theory.

As mentioned in the introduction, the counter-terms which are used for finite renormalization in curved space-times \cite{Hollands:2001nf} are local and covariant functionals $\cO$ of the field configurations $(\Phi, \Phi^\ddag)$ and the background metric $g$, which by definition are of the form

\begin{equation}
\mathcal{O}(x)= \mathcal{O}\left( g_{\mu \nu}(x), {R^\mu}_{\nu \rho \sigma}(x), \dots, \nabla_{(\mu_1 \dots} \nabla_{\mu_k)} {R^\mu}_{\nu \rho \sigma}|_x, \nabla_{(\mu_1 \dots} \nabla_{\mu_k)} \Phi |_{x}, \nabla_{(\mu_1 \dots} \nabla_{\mu_k)}\Phi^\ddag|_x \right), 
\end{equation}
where ${R^\mu}_{\nu \rho \sigma}$ is the Riemann tensor. We therefore restrict to the space $\textbf{P}(M)$ of such functionals. 
The ghost number $q$ induces a grading on the space $\textbf{P}(M)$ of all local and covariant functionals
\beq
\textbf{P}(M) = \bigoplus_{p, q} \textbf{P}^p_q(M),
\eeq
where each $\textbf{P}^p_q(M)$ is defined to be the space of functionals with ghost number $q$ and form degree $p$.
The $q$-th cohomology ring of $s$ at form degree $p$ is defined by
\begin{equation}
H_q^p (s, M):= \frac{\{ \text{ker } s: \textbf{P}^p_q(M) \rightarrow \textbf{P}^p_{q+1}(M)\}}{\{\text{im } s: \textbf{P}^p_{q-1}(M) \rightarrow \textbf{P}^p_{q}(M) \}}.
\end{equation}
The anomaly $A = \int_M a(x) dx$ is shown \cite{Hollands:2007zg} to be a formal power series in $\hbar$ 
\beq
A = A^{(m)} \hbar^m + A^{(m+1)} \hbar^{m+1} + \dots,
\eeq
whose leading order contribution $A^{(m)}$ is an element of $H_1^3 (s, M)$ of dimension $3$ and ghost number $1$. Equivalently, the local density $a^{(m)}(x)$ belongs to $H_1^3 (s|d, M)$, the \textit{cohomology ring of $s$ modulo $d$} defined by
\begin{equation}
\label{cohom-s|d}
H_q^p (s|d, M):= \frac{\{ \mathcal{O}_q^p | s \mathcal{O}^p_q= d \mathcal{O}_{q+1}^{p-1} \}}{\{  \mathcal{O}^p_q | \mathcal{O}^p_q= s \mathcal{O}^p_{q-1} +  d \mathcal{O}_{q}^{p-1} \}}.
\end{equation}  

In our formalism, we are working off-shell (where the equations of motion are not necessarily satisfied) because of the open nature of the supersymmetry algebra \eqref{deltaepsilon2}. We must, thus, consider the fields and anti-fields $(\Phi, \Phi^\ddag)$ and their derivatives 
as independent and we will refer to them as \emph{basic fields}. 

To analyse the cohomology ring \eqref{cohom-s|d} which contains potential anomalies, we first collect two well-known results in the homological perturbation theory \cite{Maggiore:1994dw, henneaux1992quantization}. 

1) Two basic fields $\Phi_1$ and $\Phi_2$ are said to form a \emph{BRST doublet} or a \emph{contractible pair}, denoted by $\{ \Phi_1, \Phi_2 \}$, if $s\Phi_1= \Phi_2$, (and hence $s \Phi_2=0)$. Then, it follows that the cohomology classes $H_q^p (s|d, M)$ are independent of the pair of fields $\{ \Phi_1, \Phi_2 \}$ forming a BRST doublet. To see this, consider for instance an $\cO$ which contains $\Phi_1$ and $\Phi_2$, $k$ times, i.e.
 \beq
(\Phi_1 \frac{\partial}{\partial \Phi_1} + \Phi_2 \frac{\partial}{\partial \Phi_2} ) \cO = k \cO.
 \eeq 
Then $\cO$ is $s$-exact, i.e. $\cO= s \cO'$ with $\cO' = \Phi_1 \frac{\partial \cO}{\partial \Phi_2}$.
 
2) Let $\mathcal{N}: \textbf{P}(M) \rightarrow \textbf{P}(M)$ be a \emph{filtration operator}, with non-negative eigenvalues, such that each element $\mathcal{O} \in\textbf{P}(M)$ and the differential $s$ have an expansion of the form
\begin{align}
\mathcal{O} &= \sum_{n \ge 0} \mathcal{O}_{n},  \\ \label{s=s_0 + ...}
s &= s_{0} + s_{1} + \dots,
\end{align}
where $\mathcal{N} \mathcal{O}_{n} := n  \mathcal{O}_{n}$. Then for a filtration $\mathcal{N}$  which commutes with the exterior derivative $d$, i.e., $[\mathcal{N}, d]=0$, we have
\begin{enumerate}
\item $s_{0}^2=0$,
\item $H_q^p (s|d, M)$ is isomorphic to a subgroup of $H_q^p (s_{0}|d, M)$.
\end{enumerate}
The first result is a trivial consequence of expanding $s^2=0$. For the second one, consider an $\cO$ with $s \cO=0$. Using the expansion \eqref{s=s_0 + ...}, this means
\begin{align*}
s_{0} \cO_{0} &=0, \\
s_{1} \cO_{0} + s_{0} \cO_{1} &=0, \\
s_{2} \cO_{0} + s_{1} \cO_{1} + s_{0} \cO_{2} &=0, \\
\dots.
\end{align*}
This means $\cO_{(0)}$ is in $s_{0}$-cohomology. One can show (\cite{Maggiore:1994dw} Proposition 5.6) that, in fact, the map $\rho (\cO) = \cO_{0}$ provides the isomorphism mentioned above.
\subsection{Triviality of cohomology rings $H_1^3 (s|d, M)$ and $H_1^3 (s, M)$} \label{BRST-trivial}
In this section, we prove that the cohomology rings $H_1^3 (s|d, M)$ and $H_1^3 (s, M)$ at dimension $3$ are trivial.
The proof rests on a suitable filtration which decomposes the BRST differential as in \eqref{s=s_0 + ...}. With respect to $s_{0}$, part of the fields and their derivatives form doublets and hence do not belong to  $H_1^3 (s_{0}|d, M)$. We then calculate this cohomology in two steps:
\begin{enumerate}
\item[(1)] We write the most general element $\cO \in \textbf{P}_1^3(M)$ with dimension $3$, as a linear combination of all possible local-covariant functionals with form degree $3$, and ghost number $1$, made out of the remaining fields (which do not form doublets),
\item[(2)] By applying $s_{0}$, we show that all the coefficients of this linear combination have to vanish, in order that $A \in H_1^3(s_{0}|d, M)$.
\end{enumerate}
Since $H_1^3 (s|d, M) \subset H_1^3 (s_{0}|d, M)$, it follows that $H_1^3 (s|d, M)$, hence $H_1^3 (s, M)$, are trivial.
As filtration operator, we choose
\beq
\mathcal{N}= \int_M n_\Phi \Phi \frac{\delta}{\delta \Phi} + n_{{\Phi}^\ddag} \Phi^\ddag \frac{\delta}{\delta \Phi^\ddag},
\eeq
which counts the number of fields and anti-fields $(\Phi, \Phi^\ddag)$ with weights $(n_\Phi, n_{\Phi^\ddag})$ which are given in the Table \ref{table3}. The filtration is made in such a way that 
\beq
[\mathcal{N}, d]=0 = [\mathcal{N}, \nabla_\mu],
\eeq 
therefore $n_{d\Phi} = n_{\nabla_\mu \Phi}= n_\Phi$.
\begin{table}
\begin{center}
    \begin{tabular}{   | l | l | l | l | l | l | l | l | l | l | p{0.4 cm}  | }
    \hline
    $\Phi$&  $(A_\mu, \hat{A}_\mu)$ &$\mathbb{X}_A$ &  $\Psi_A$ &$(c, \hat{c})$ & $X_\mu$ &${\alpha^I}_J$ &$\epsilon^I$ & $(\bar{c}, \bar{\hat{c}})$  & $(B, \hat{B})$ \\ \hline 
    $n_\Phi$ & $2$ &  $3$ & $2$ &$2$ &$ 2$  &$2$ & $1$& $2$ & $2$\\ \hline \hline
    $\Phi^\ddag$& $(A_\mu^\ddag, \hat{A}_\mu^\ddag)$  & $\mathbb{X}_A^\ddag$ & $\Psi_A^\ddag$ & $(c^\ddag, \hat{c}^\ddag)$&  $X^\ddag_\mu$ &${\alpha^{\ddag I}}_J$ &$\epsilon^{\ddag I}$ &   $(\bar{c}^\ddag, \bar{\hat{c}}^\ddag)$ & $(B^\ddag, \hat{B}^\ddag)$\\ \hline 
    $n_{\Phi^\ddag}$ & $3$  & $1$ & $2$ & $2$ & $1$ & $3$ &$2$ & $2$ & $2$\\ \hline 
    \end{tabular}
    \caption{\label{table3} Fields, anti-fields and their weights $n_{\Phi}, n_{\Phi^\ddag}$.}
\end{center}
\end{table}
The zeroth order part, therefore, satisfies $s^2_{0}=0$ and  
\beq
s_{0} \nabla_\mu - \nabla_\mu s_{0} =0= s_{0} d + d s_{0}.
\eeq
The fields and anti-fields with non-vanishing $s_{0}$ transformations are
\begin{equations}[hats0]
&(s_{0} A_\mu,  s_{0}\hat{A}_\mu) = (\nabla_\mu c, \nabla_\mu \hat{c}), \\ 
& s_{0} \mathbb{X}_A = i \Gamma^I_{AB} \bar{\epsilon}^I \Psi^B,\\ 
& s_{0} X_\mu = \xi_\mu,\\ 
&s_{0} {\alpha^I}_J =  \frac{2}{3} i\bar{\epsilon}^I {\not}\nabla\epsilon_J,\\
& (s_{0}\bar{c}, s_{0}\bar{\hat{c}}) = (B, \hat{B}),\\
& (s_{0} c^\ddag,  s_{0}\hat{c}^\ddag) = (\Box c, \Box \bar{c} ),\\
&(s_{0} A_\mu^\ddag,  s_{0}\hat{A}_\mu^\ddag) = ( 2 \epsilon_{\mu \lambda \rho} \nabla^\lambda A^\rho, 2 \epsilon_{\mu \lambda \rho} \nabla^\lambda \hat{A}^\rho),\\ 
& s_{0} \Psi_A^\ddag = {\not} \nabla \Psi_A,\\ 
&(s_{0} B^\ddag, s_{0} \hat{B}^\ddag) = (B, \hat{B}),\\ 
& s_{0} \epsilon^{\ddag I } = - i \bar{\epsilon}^I \gamma_\mu X^{\ddag \mu}, \\ 
& s_{0} {\alpha^{\ddag I}}_J =  \epsilon^{\ddag I} \epsilon_J.
\end{equations}
We observe that the following fields form $s_{0}$-doublets,
\begin{align*}
&\{ A_\mu, \nabla_\mu c \}, \hspace{3 mm} \{ \hat{A}_\mu, \nabla_\mu \hat{c}\}, \hspace{3 mm} \{ \bar{c}, B\}, \hspace{3 mm}\{ \bar{\hat{c}}, \hat{B}\}, \hspace{3 mm}  \{ \Psi_A^\ddag,{\not} \nabla \bar{\Psi}_A\},\\
  \{ A_\mu^\ddag,  \epsilon_{\mu \nu \rho}&\nabla^\nu A^\rho \}, \hspace{1 mm} \{ \hat{A}_\mu^\ddag, \epsilon_{\mu \nu \rho}\nabla^\nu \hat{A}^\rho \},  \hspace{1 mm}  \{ c^\ddag, \Box \bar{c}\} , \hspace{1 mm} \{ \hat{c}^\ddag, \Box \bar{\hat{c}}\}, \hspace{1 mm} \{ B^{\ddag}, B\}, \hspace{1 mm} \{ \hat{B}^{\ddag}, \hat{B}\}.
\end{align*}
Also, since both $\mathcal{N}$ and $s_{0}$ (anti-)commute with $\nabla_\mu$ and $d$, all derivatives of each doublet form again new doublets, e.g. $\{ \nabla_\nu A_\mu, \nabla_\nu \nabla_\mu c \}$. Hence, such fields do not belong to the $s_{0}$-cohomology.

A general element  $\cO \in H_1^3(s_0|d, M)$  is a 3-form and ghost number $+1$, and satisfies $s_{0} \cO=d \theta$, for some $\theta \in \textbf{P}_2^2(M)$.  Here we are only interested in the anomaly $a(x)$ which is of dimension $3$. Let us denote the Hodge dual of $a$ by $\mathcal{A}$ which is a scalar. To have ghost number $+1$, $\mathcal{A}$ has to either
\begin{itemize}
\item be linear in ghosts $(c, \hat{c}), \epsilon, X, \alpha$ with ghost number $+1$, which we denote by $\mathcal{A}(c, \hat{c})$, $\mathcal{A}(\epsilon)$, $\mathcal{A}(X)$, $\mathcal{A}(\alpha)$ respectively, or
\item depend on $\mathbb{X}_A^\ddag, \epsilon^\ddag, \alpha^\ddag, X^\ddag$ with negative ghost numbers, and enough number of other ghosts to make a combination with ghost number $+1$. We denote them by $\mathcal{A}(\mathbb{X}^\ddag)$, $\mathcal{A}(\epsilon^\ddag)$, $\mathcal{A}(\alpha^\ddag)$, $\mathcal{A}(X^\ddag)$, respectively.
\end{itemize}
The most general manifestly gauge and $\mathfrak{so}(6)$-invariant form that these terms can take are
\begin{align*}
\mathcal{A}(c, \hat{c})&=  \text{Tr} \left(c  (a_1  \mathbb{X}_A \mathbb{X}^A \mathbb{X}_B \mathbb{X}^B \mathbb{X}_C \mathbb{X}^C  + a_2 \mathbb{X}_A \mathbb{X}^B \mathbb{X}_C \mathbb{X}^A \mathbb{X}_B \mathbb{X}^C   + a_3 \mathbb{X}_A \mathbb{X}^B \mathbb{X}_B \mathbb{X}^A \mathbb{X}_C \mathbb{X}^C \right. \\
& \quad \left. +   a_4 \epsilon_{ABCD}\bar{\Psi}^A \mathbb{X}^B \Psi^C \mathbb{X}^D+ a_5 \bar{\Psi}^A \Psi_A \mathbb{X}_B \mathbb{X}^B + a_6 \bar{\Psi}^A \Psi_B \mathbb{X}_A \mathbb{X}^B) \right)\\
&\quad  +  \text{Tr} \left( {\hat{c}}  (\hat{a}_1 \mathbb{X}^A \mathbb{X}_A \mathbb{X}^B \mathbb{X}_B \mathbb{X}^C \mathbb{X}_C  + \hat{a}_2  \mathbb{X}^A \mathbb{X}_B \mathbb{X}^C \mathbb{X}_A \mathbb{X}^B \mathbb{X}_C + \hat{a}_3  \mathbb{X}^A \mathbb{X}_B \mathbb{X}^B \mathbb{X}_A \mathbb{X}^C \mathbb{X}_C  \right. \\
& \quad  +  \left. \hat{a}_4 \epsilon^{ABCD}\bar{\Psi}_A \mathbb{X}_B \Psi_C \mathbb{X}_D +\hat{a}_5 \bar{\Psi}_A \Psi^A \mathbb{X}^B \mathbb{X}_B +  \hat{a}_6 \bar{\Psi}_A \Psi^B \mathbb{X}^A \mathbb{X}_B) \right), \\
& \quad +  a_{7} R \text{Tr} \left(  c \mathbb{X}_A \mathbb{X}^A  \right) +  \hat{a}_{7}  R \text{Tr}\left( \hat{c} \mathbb{X}^A \mathbb{X}_A \right)\\
\mathcal{A}(\epsilon) &=   \tilde{\Gamma}^I_{EF} \bar{\epsilon}^I  \text{Tr} \left( \Psi^E  \mathbb{X}^F  (  b_1  \mathbb{X}^A \mathbb{X}_A \mathbb{X}_B \mathbb{X}^B+ b_2 \mathbb{X}^A \mathbb{X}_B \mathbb{X}_A \mathbb{X}^B) \right)\\
& \quad +  {\Gamma}^{IEF} \bar{\epsilon}^I \text{Tr}\left( \Psi_E  \mathbb{X}_F  ( \hat{b}_1  \mathbb{X}_A \mathbb{X}^A \mathbb{X}^B \mathbb{X}_B + \hat{b}_2 \mathbb{X}_A \mathbb{X}^B \mathbb{X}^A \mathbb{X}_B) \right)\\
&\quad  + b_3 \tilde{\Gamma}^I_{AB} \bar{\epsilon}^I \gamma^\mu \text{Tr} (\Psi^{A}  \mathbb{X}^B \mathbb{X}_C \nabla_\mu \mathbb{X}^C)  
+ \hat{b}_3  {\Gamma}^{IAB} \bar{\epsilon}^I \gamma^\mu \text{Tr}(\Psi_{A}  \mathbb{X}_B \mathbb{X}^C \nabla_\mu \mathbb{X}_C)\\
& \quad + b_4  \tilde{\Gamma}^I_{AB} \bar{\epsilon}^I  \gamma^\mu \gamma^\nu  \text{Tr}\left( \Psi^{A}  \nabla_\mu \nabla_\nu  \mathbb{X}^B \right) + \hat{b}_4  {\Gamma}^{IAB} \bar{\epsilon}^I  \gamma^\mu \gamma^\nu  \text{Tr} \left(\Psi_A  \nabla_\mu \nabla_\nu  \mathbb{X}_B \right)\\
&\quad  + b_5   \tilde{\Gamma}^I_{AB}  \text{Tr} \left(\nabla^\mu (\bar{\epsilon}^I  \Psi^{A})  \nabla_\mu  \mathbb{X}^B \right) + \hat{b}_5  {\Gamma}^{IAB}  \text{Tr} \left(\nabla^\mu (\bar{\epsilon}^I  \Psi_A)  \nabla_\mu  \mathbb{X}_B \right) \\
&  \quad + b_6 \tilde{\Gamma}^I_{AB}   {\not}\nabla \bar{\epsilon}^I  \text{Tr} \left( \Psi^{A}  \mathbb{X}^B \mathbb{X}_C \mathbb{X}^C\right) + 
\hat{b}_6  {\Gamma}^{IAB}   {\not}\nabla \bar{\epsilon}^I  \text{Tr} \left(\Psi_{A}  \mathbb{X}_B \mathbb{X}^C \mathbb{X}_C \right)\\
&\quad  + b_7 R \tilde{\Gamma}^I_{AB} \bar{\epsilon}^I \text{Tr} \left( \Psi^{A}  \mathbb{X}^B \right) + \hat{b}_7 R  {\Gamma}^{IAB} \bar{\epsilon}^I \text{Tr} \left(\Psi_A  \mathbb{X}_B \right) \\
& \quad + b_8  \tilde{\Gamma}^I_{AB} \bar{\epsilon}^I \text{Tr} \left( \nabla^2 \Psi^{A}  \mathbb{X}^B \right) + \hat{b}_8   {\Gamma}^{IAB} \bar{\epsilon}^I  \text{Tr} \left(\nabla^2 \Psi_A  \mathbb{X}_B \right), \\
\mathcal{A}(X) & = \sigma_X \text{Tr} \left(  k_1  \mathbb{X}_A \mathbb{X}^A \mathbb{X}_B \mathbb{X}^B \mathbb{X}_C \mathbb{X}^C  + k_2 \mathbb{X}_A \mathbb{X}^B \mathbb{X}_C \mathbb{X}^A \mathbb{X}_B \mathbb{X}^C   + k_3 \mathbb{X}_A \mathbb{X}^B \mathbb{X}_B \mathbb{X}^A \mathbb{X}_C \mathbb{X}^C \right. \\
& \quad \left. +   k_4 \epsilon_{ABCD}\bar{\Psi}^A \mathbb{X}^B \Psi^C \mathbb{X}^D+ k_5 \bar{\Psi}^A \Psi_A \mathbb{X}_B \mathbb{X}^B + k_6 \bar{\Psi}^A \Psi_B \mathbb{X}_A \mathbb{X}^B \right)\\
&\quad  + \sigma_X  \text{Tr} \left(   \hat{k}_1 \mathbb{X}^A \mathbb{X}_A \mathbb{X}^B \mathbb{X}_B \mathbb{X}^C \mathbb{X}_C  + \hat{k}_2  \mathbb{X}^A \mathbb{X}_B \mathbb{X}^C \mathbb{X}_A \mathbb{X}^B \mathbb{X}_C + \hat{k}_3  \mathbb{X}^A \mathbb{X}_B \mathbb{X}^B \mathbb{X}_A \mathbb{X}^C \mathbb{X}_C  \right. \\
& \quad  +  \left. \hat{k}_4 \epsilon^{ABCD}\bar{\Psi}_A \mathbb{X}_B \Psi_C \mathbb{X}_D +\hat{k}_5 \bar{\Psi}_A \Psi^A \mathbb{X}^B \mathbb{X}_B +  \hat{k}_6 \bar{\Psi}_A \Psi^B \mathbb{X}^A \mathbb{X}_B \right),\\
\mathcal{A}(\alpha) & = d_1 {\alpha^I}_J  \Gamma^{I AB} \tilde{\Gamma}^J_{CD}    \text{Tr} (\mathbb{X}_A \mathbb{X}^C \mathbb{X}_B \mathbb{X}^D \mathbb{X}_E  \Phi^{E})
 +  d_2{\alpha^I}_J  \Gamma^{I AB} \tilde{\Gamma}^J_{CD} \text{Tr} (\bar{\Psi}^{C} \Psi_A \mathbb{X}_B \mathbb{X}^D) \\
 & \quad + \hat{d}_1 {\alpha^I}_J \Gamma^{I AB} \tilde{\Gamma}^J_{CD}    \text{Tr}( \mathbb{X}^C \mathbb{X}_B \mathbb{X}^D \mathbb{X}_A \mathbb{X}^E  \mathbb{X}_E)
 +  \hat{d}_2 {\alpha^I}_J  \Gamma^{I AB} \tilde{\Gamma}^J_{CD} \text{Tr} (\bar{\Psi}_{A} \Psi^C \mathbb{X}^D \mathbb{X}_B),\\
\mathcal{A}(\mathbb{X}^\ddag) & =  \text{Tr}\left( e_1 c^2 \mathbb{X}_A \mathbb{X}^{\ddag A} + \hat{e}_1 \hat{c}^2 \mathbb{X}^A \mathbb{X}_A^{\ddag} \right)  +  e_2 \text{Tr}(\mathbb{X}_A \mathbb{X}^{\ddag A})  {\alpha^I}_J {\alpha^J}_I +   \hat{e}_2 \text{Tr}(\mathbb{X}^A \mathbb{X}_A^{\ddag})  {\alpha^I}_J {\alpha^J}_I,  \\
\mathcal{A}(\epsilon^\ddag) &=  f_1\epsilon^\ddag_I \epsilon^I \bar{\epsilon}_J \epsilon^J \text{Tr}\left( \mathbb{X}_A \mathbb{X}^A \right) + f_2 \epsilon_I^\ddag \gamma^\mu \epsilon^I X_\mu \text{Tr} \left( c \mathbb{X}_A \mathbb{X}^A \right) + f_3 \epsilon^\ddag_I \epsilon^I c^2 +  f_4 \epsilon^\ddag_I \epsilon^I  {\alpha^I}_J {\alpha^J}_I\\
& \quad +\hat{f}_1\epsilon^\ddag_I \epsilon^I \bar{\epsilon}_J \epsilon^J \text{Tr}\left( \mathbb{X}^A \mathbb{X}_A \right) 
 + \hat{f}_2 \epsilon_I^\ddag \gamma^\mu \epsilon^I X_\mu \text{Tr} \left( \hat{c} \mathbb{X}^A \mathbb{X}_A \right) + \hat{f}_3 \epsilon^\ddag_I \epsilon^I \hat{c}^2,\\
\mathcal{A}(\alpha^\ddag) &= g_1 {\alpha^{\ddag I}}_J \bar{\epsilon}^J \epsilon_I \text{Tr} \left(c \mathbb{X}_A \mathbb{X}^A \right)  + \hat{g}_1 {\alpha^{\ddag I}}_J \bar{\epsilon}^J \epsilon_I \text{Tr} \left(c \mathbb{X}^A \mathbb{X}_A\right) \\
& \quad +{\alpha^{\ddag I}}_J  \bar{\epsilon}^J \gamma^\mu \epsilon_I X_\mu  \text{Tr}\left(g_2 \mathbb{X}^A \mathbb{X}_A \mathbb{X}_B \mathbb{X}^B+ g_3 \mathbb{X}^A \mathbb{X}_B \mathbb{X}_A \mathbb{X}^B\right)\\
&\quad  +  {\alpha^{\ddag I}}_J  \bar{\epsilon}^J \gamma^\mu \epsilon_I X_\mu  \text{Tr}\left(\hat{g}_2  \mathbb{X}_A \mathbb{X}^A \mathbb{X}^B \mathbb{X}_B + \hat{g}_3 \mathbb{X}_A \mathbb{X}^B \mathbb{X}^A \mathbb{X}_B \right) \\
& \quad + g_5{\alpha^{\ddag I}}_J {\not}\nabla \bar{\epsilon}^J \epsilon_I c + \hat{g}_5{\alpha^{\ddag I}}_J {\not}\nabla \bar{\epsilon}^J \epsilon_I \hat{c},\\
\mathcal{A}(X^\ddag) &= h_1 X^{\mu \ddag} X_\mu \text{Tr } c^2 + \hat{h}_1 X^{\mu \ddag} X_\mu \text{Tr } \hat{c}^2  + h_2 X^{\mu \ddag} X_\mu {\alpha^I}_J {\alpha^J}_I + h_3 X^{\mu \ddag} X_\mu \bar{\epsilon}^I \epsilon_I,
\end{align*}
where $a_i, \hat{a}_i, b_i, \hat{b}_i, k_i, \hat{k}_i, d_i, \hat{d}_i, e_i, \hat{e}_i, f_i, \hat{f}_i, g_i, \hat{g}_i, h_i, \hat{h}_i $ are constants, and $R$ is the scalar curvature of the background spacetime.
Using the relatively simple form of $s_{0}$ given in \eqref{hats0}, one can straightforwardly check that the above candidate element is in $H^3_1(s_{0}|d, M)$ only if all constants vanish. For instance, if we apply $s_{0}$ on $\mathcal{A}(c, \hat{c})$, we observe that since $s_{0} \mathbb{X}_A $ contains a superconformal ghost $\epsilon^I$, $s_{0} \mathcal{A}(c, \hat{c})$ will necessarily contain $(c, \hat{c}) \epsilon^I$. But these ghosts cannot appear in this combination as the $s_{0}$ transformation of any other terms in the above expression. Thus, to be in the cohomology, all the coefficients $a_i, \hat{a}_i$ must vanish. Similarly, applying $s_{0}$ on $\mathcal{A}(\epsilon)$ and $\mathcal{A}(X)$ will result in ghosts in the combination $\bar{\epsilon}^I \epsilon^J$ and $\sigma_X \epsilon^I$ respectively which, however, cannot be generated as the $s_0$ transformation of any of the other terms, hence $a_i, \hat{a}_i$ and $k_i, \hat{k}_i$ must vanish as well. 
\section*{Acknowledgement}
This work is part of the author's PhD dissertation. I gratefully acknowledge financial
support by the Max Planck Institute for Mathematics in the Sciences and its International
Max Planck Research School (IMPRS). I am very grateful to my supervisor Stefan Hollands for suggesting me the subject, for fruitful discussions  and a critical reading of the manuscript.
\appendix 
\section{BRST transformations of anti-fields}\label{BRSTantifields}
In this appendix, we give the $s$ transformation of anti-fields with a non-trivial $s_{0}$ transformation used in Section \ref{BRST-trivial}.
\begin{align*}
& s A^{\mu\ddag} =  \xi_\nu (F^{\mu \nu} - \epsilon^{\mu\nu \rho} J_\rho) + [A^{\mu \ddag}, c] + (\mathcal{L}_X - 3 \sigma_X) A^{ \mu \ddag} -\frac{1}{2} \xi^\mu c^\ddag  ,\\
& s \Psi_A^{\ddag} =  {\not} D \bar{\Psi}_A - 2 \epsilon_{ABCD} \mathbb{X}^B \bar{\Psi}^C \mathbb{X}^D - \mathbb{X}^B\mathbb{X}_B \bar{\Psi}_A + \bar{\Psi}_A \mathbb{X}_B \mathbb{X}^B - 2 \bar{\Psi}_B \mathbb{X}_A \mathbb{X}^B + 2 \mathbb{X}^B \mathbb{X}_A \bar{\Psi}_B\\
& \quad  -i c\Psi^\ddag_A + i \Psi^\ddag_A \hat{c}  + ({\Gamma^I}_J)_B^A {\alpha^J}_I \Psi^{\ddag}_B +  ( \mathcal{L}_X { - 2}\sigma_X) \Psi^\ddag_A +  \Gamma^I_{AB}\bar{\epsilon}^I \gamma_\mu \mathbb{X}^B (A^{\ddag \mu} + \hat{A}^{\ddag \mu}) +  i \Gamma^I_{BA} \bar{\epsilon}^I \mathbb{X}^{\ddag B}   \\
&s c^\ddag = - D_\mu \nabla^\mu \bar{c} + i[A_\mu, A^{\ddag \mu}]  - i \mathbb{X}_A \mathbb{X}^{\ddag A} -  i \Psi_A \Psi^{\ddag A} - [c, c^\ddag] - (\mathcal{L}_X -3 \sigma_X) c^\ddag\\
&s {\alpha^{\ddag I}}_J =  ({\Gamma^I}_J)_A^B \mathbb{X}_B \mathbb{X}^{\ddag A}+  ({\Gamma^I}_J)_A^B \Psi_B \Psi^{A \ddag} -{\alpha^I}_K {\alpha^{\ddag K}}_J + \epsilon^{I \ddag} \epsilon_J  \\ \nonumber
& s \epsilon^{\ddag I}= \tilde{\Gamma}^{IAB} \mathbb{X}_B \bar{\Psi}_A {\not} \nabla \bar{c} - i \epsilon_{\mu \nu \rho} \bar{\epsilon}^I \gamma^\nu A^{\ddag \rho} \nabla^\mu \bar{c} - \tilde{\Gamma}^{IAB} \mathbb{X}_B \bar{\Psi}_A \gamma_\mu (A^{\ddag \mu} + \hat{A}^{\ddag \mu}) \\
&\quad  + i \epsilon_{\mu \nu \rho} \bar{\epsilon}^I \gamma^\nu ( A^{\ddag \rho} A^{\ddag \mu} +  \hat{A}^{\ddag \rho} \hat{A}^{\ddag \mu}) + i \Gamma^I_{AB} \bar{\Psi}^B \mathbb{X}^{\ddag A} - \frac{1}{3} \Gamma^I_{AB} {\not}\nabla( \mathbb{X}^B \Psi^{\ddag A}) \\ \nonumber
&\quad   + \Gamma^I_{AB} {\not}D \mathbb{X}^B \Psi^{\ddag A} + {\Gamma}^I_{AB} (\mathbb{X}^C \mathbb{X}_C \mathbb{X}^B - \mathbb{X}^B \mathbb{X}_C \mathbb{X}^C)\Psi^{\ddag A} - 2 {\Gamma}^I_{BC} \mathbb{X}^B \mathbb{X}_A \mathbb{X}^C \epsilon^I     -\frac{1}{2} i  \bar{\epsilon}^I \gamma_\mu \bar{\Psi}_A^{\ddag} \Psi^{\ddag A}\gamma^\mu , \\
& \quad  -  {\frac{1}{2}( i \bar{\epsilon}^I \gamma^\mu (A_\mu c^\ddag + \hat{A}_\mu \hat{c}^\ddag)  - (\bar{\epsilon}^J({\Gamma^I}_J)_D^C \mathbb{X}^D \mathbb{X}_C)(c^\ddag + \hat{c}^\ddag) ) } + \frac{2}{3}i {\not}\nabla ({\alpha^{\ddag I}}_J \bar{\epsilon}^J) - i \bar{\epsilon}^I \gamma_\mu X^{\ddag \mu} + {\alpha^I}_J \epsilon^{\ddag J}\\ \nonumber
& s B^\ddag = - \nabla_\mu A^\mu + \alpha B + \alpha \sigma_X \bar{c} + (\mathcal{L}_X - 2 \sigma_X)B^\ddag + c^\ddag.
\end{align*}

\addcontentsline{toc}{section}{References}
\bibliography{literature}

\end{document}